\documentclass[3p,twocolumn]{elsarticle}

\usepackage[textsize=tiny]{todonotes}

\usepackage{dcolumn}
\usepackage{amsbsy}
\usepackage{amsmath}

\usepackage{amssymb}
\usepackage{bm}
\usepackage{braket}
\usepackage{caption}
\usepackage{color}
\usepackage{graphicx}

\usepackage{cleveref}
\usepackage{subfigure}
\usepackage{siunitx}

\usepackage{comment}

\newcommand{\ba}{\bm{a}}

\newcommand{\bc}{\bm{c}}

\newcommand{\be}{\bm{e}}

\newcommand{\bv}{\bm{v}}

\newcommand{\bzero}{\mathbf{0}}

\newcommand{\bkappa}{\boldsymbol{\kappa}}

\begin{document}
\title{Tunable wave propagation by varying prestrain in tensegrity-based periodic media}
\author{Raj Kumar Pal$^{a}$, Massimo Ruzzene$^{a,b}$ and Julian J. Rimoli$^{a}$\\ 
\small $^a$ School of Aerospace Engineering, Georgia Institute of Technology, Atlanta GA 30332 \\
\small $^b$ School of Mechanical Engineering, Georgia Institute of Technology, Atlanta GA 30332}




\begin{abstract}
This paper investigates the dynamic properties of one, two and three-dimensional tensegrity-based periodic structures introduced 
in~\cite{rimoli2017mechanical}, which are here termed as tensegrity beams, plates and solids, respectively. 
We study their linear wave propagation properties
and show that in each case, these properties can be significantly altered by the prestrain in the cables. As the prestrain is varied, 
we observe jumps in the wave velocities at two critical prestrain values, which define transitions between the three distinct phases of these
structural assemblies. At low cable prestrains, the wave speeds are zero as the lattices have zero effective stiffness. At moderate prestrains, 
the wave speed is nonzero and finally, at prestrain levels where the bars buckle, the wave speed decreases to a lower value. Dispersion analysis on these beams, plates and solids reveal unique properties such as very low wave velocities 
compared to their constituent material and the existence of flat bands at low frequencies.  Furthermore, we find that 
shear waves travel faster than longitudinal waves in tensegrity solids in a range of cable prestrains.  
Finally, we verify the key observations through detailed numerical simulations on finite tensegrity solids. 
\end{abstract}
\maketitle

\section{Introduction}
Artificially engineered lattices have been the subject of extensive research over the past decade 
due to their unique static and dynamic properties~\cite{hussein2014dynamics,christensen2015vibrant}. 
Examples include pentamode lattices with zero shear modulus~\cite{buckmann2014elasto,wang2017composite}, 
auxetic lattices with negative Poisson's ratio~\cite{babaee20133d,lakes2017negative} 
and  hexagonal lattices featuring frequency bandgaps~\cite{ruzzene2003wave,tallarico2017tilted}. 
These lattices fall under the category of metamaterials and metastructures that have 
potential for unprecedented control over the effective structural properties
through the careful design of their microstructure. 
In recent years, there has been an increased focus on developing structures and lattices whose mechanical properties
can be significantly altered by applying external fields. Examples include elastomers with holes in periodic arrangements~\cite{bertoldi2017harnessing}, deformable hexagonal lattices~\cite{pal2016effect}, origami~\cite{filipov2015origami} and tensegrity based structures~\cite{rimoli2017mechanical,liu2017programmable}, among others. 

Tensegrity structures are meta-structures composed of an assembly of prestressed bars and cables,  
connected in such a way that the cables (bars) are in tension (compression) and the structure is in equilibrium 
without any external force~\cite{richard1962tensile,pugh1976introduction}. 
Simple examples of tensegrity structures include an archery bow and the human bone tendon arrangement~\cite{skelton2009tensegrity}. 
Their structural integrity arises as a consequence of the member prestrain. 
A key property of these structures which makes them distinct from conventional 
load carrying ones is that their compression carrying members (bars) are isolated while the tensile load carrying members (cables) are all connected.  
Tensegrity lattices offer unique possibilities for tunable static~\cite{rimoli2017mechanical} 
and dynamic properties~\cite{amendola2018tuning} by varying the cable prestrain.
Potential applications range from
deployable structures~\cite{skelton2009tensegrity} to 
robotics~\cite{paul2005gait} and impact mitigation structures~\cite{rimoli2017reduced}. 

Developing a periodic tensegrity lattice is a nontrivial task since most of the tensegrity structures lack
translation symmetry and cannot tessellate a $3D$ domain. There have been a few notable 
examples in the past decade for developing 
tensegrity-based beams and plates~\cite{skelton2009tensegrity,fraternali2014multiscale}. 
However, these designs have a continuous path of bars spanning the entire structure 
and do not share the unique topological properties of tensegrity structures like localized span of compressive elements. 
Moreover they do not easily extend to $3D$ tensegrity solids. 
Recently, Rimoli and Pal~\cite{rimoli2017mechanical} developed a unit cell for a $3D$ periodic tensegrity lattice 
by exploiting the symmetries of a truncated octahedron tensegrity sphere. 
This lattice satisfies all the conditions of a tensegrity structure in the sense that only the tensile members are path connected to one another. 

In this letter, we investigate the dynamic properties of tensegrity based $1D$, $2D$ and $3D$ periodic structures 
introduced in~\cite{rimoli2017mechanical}, which are here termed as tensegrity beams, plates and solids, respectively. 
We study their linear wave properties using dispersion analysis and show how these properties can be significantly altered 
by varying the cable prestrain. 
We model the bars as elastic or deformable systems in a geometrically nonlinear elasticity 
framework in contrast to prior 
investigations~\cite{connelly1995globally,kahla2000nonlinear,zhang2015geometrically,vassart1999multiparametered} 
that treat the bars and cables as perfectly rigid or implicitly assume that cables (bars) only experience tension (compression).
As the prestrain is varied, we observe jumps in wavespeeds at two critical prestrain values and draw parallels with phase transitions in
condensed matter systems. Our dispersion analyses also reveal unique properties like faster shear waves than longitudinal waves for a range of cable prestrain in tensegrity solids. 

The outline of this letter is as follows: In Sec.~\ref{TheorySec} we briefly describe the lattice unit cell and
our modeling methodology. Section~\ref{dispSec} presents dispersion diagrams and wave speeds 
at low frequencies under a range of cable prestrains, illustrating the unique dynamics of these lattices. 
Finally, the conclusions are summarized in Sec.~\ref{concSec}.

\section{Lattice description and modeling approach}\label{TheorySec}

\begin{figure*}
\centering
\hspace{0.35in}
\subfigure[]{
\includegraphics[width=0.164\textwidth,trim={0 -1.5cm 0 0},clip]{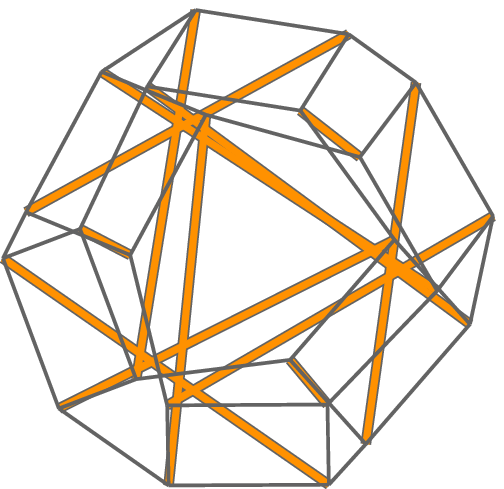}
\label{Tsphere}
}
\hspace{1.0in}
\subfigure[]{
\includegraphics[width=0.3\textwidth]{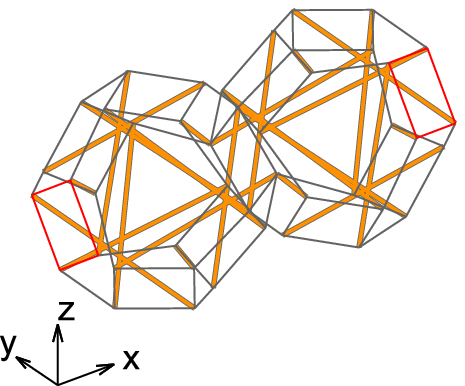}
\label{Tbeam}
}\\
\subfigure[]{
\includegraphics[width=0.4\textwidth]{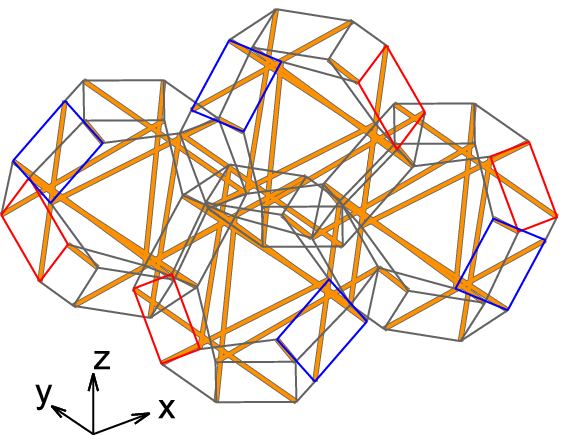}
\label{Tplate}
}
\subfigure[]{
\includegraphics[width=0.4\textwidth]{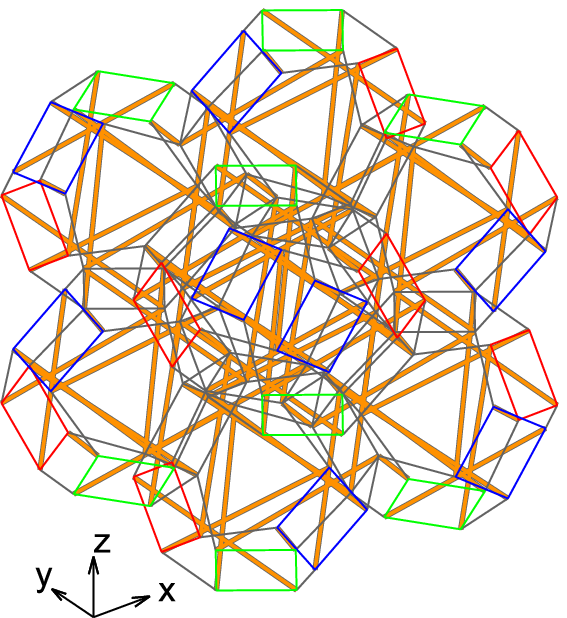}
\label{Tsolid}
}
\caption{(a) Tensegrity sphere with a truncated octahedron geometry. Unit cells for a tensegrity beam (b), plate (c)
and solid (d), obtained by successive reflection operations to get translation symmetry
and preserve discontinuity in compression. After~\cite{rimoli2017mechanical}.} 
\label{Tlattice}
\end{figure*}

The tensegrity structure employed as the building block for the tensegrity beams, plates and solids is described in detail 
in~\cite{rimoli2017mechanical} and depicted in Fig.~\ref{Tsphere}. Its 
nodes lie at the vertices of a truncated octahedron and it comprises
of $12$ bars (black, dark lines) in the interior and $36$ cables (red, light lines) 
along the edges of the octahedron. The bars and cables intersect at nodes, which are modeled
as perfect pin joints. 
This structure lacks translation symmetry as the two square facets on opposite ends 
are twisted about their normal axis relative to each other. 
According to~\cite{rimoli2017mechanical}, a unit cell with translation symmetry can obtained by 
performing a sequence of reflection operations. These operations lead to a unit cell for a tensegrity beam 
(periodic along $x$ in Fig.~\ref{Tbeam}), 
plate (periodic along $x$ and $y$ in Fig.~\ref{Tplate}) and solid (periodic along $x$, $y$ and $z$ in 
Fig.~\ref{Tsolid}), satisfying the required translation symmetries
along the lattice vector directions. 
The tensegrity lattice also has a $C_4$ rotational symmetry about the axes passing through the center of the unit cell
and coinciding with the lattice vector directions $\ba_k$. 
Note that each unit cell in isolation has $C_2$ symmetry, while the $\pi/4$ rotational symmetry 
is obtained by gliding $|\ba_k|/2$ along this axis, resulting in an infinite lattice having $C_4$ symmetry. 

A key feature of 
tensegrity structures is that the level of tension (compression) can be varied by varying the member prestrain, 
characterized here by a cable prestrain parameter $\lambda$. 
It is defined with respect to our canonical unit cell, which is a purely geometric entity 
as illustrated in Fig.~\ref{Tlattice}, with the unique property that all bar (cable) lengths are equal. 
$\lambda$ is the ratio of the undeformed cable length to the length of the cable 
segment in this canonical unit cell. 
We assume here that all the cables have identical prestrain parameter $\lambda$ and that the bar lengths
are identical to that in the canonical unit cell. 
When the bars and cables (with $\lambda < 1$) are assembled, 
the cables get stretched and the bars get compressed until an equilibrium configuration is reached. This equilibrium
configuration is determined by using  a Newton-Raphson procedure~\cite{pal2016continuum,rimoli2017mechanical} 
and is taken to be the reference
configuration for the lattice. As $\lambda$ varies, the reference configuration thus 
changes. The dispersion and wave propagation analyses are conducted on this reference configuration. 
The lattice vectors in each periodic direction are determined by the difference of position vectors between the 
corresponding nodes on opposite square facets in Fig.~\ref{Tlattice}. 
Since the cable prestrains are equal, the lattice vectors are of equal magnitude and are orthogonal due to 
symmetry. 

The bars and cables are modeled as beams made of linear elastic materials, capable of undergoing large
deformations. We use an equivalent spring mass model for each beam, where the spring stiffness and masses are chosen 
to match the axial stiffness, total mass and mass moment of inertia, and buckling load of the continuum 
beam. Each beam comprises of $4$ point masses, with each mass having three translational degrees of freedom. 
The mass at each node is simply the sum of  the end mass of all the bars and cables meeting there. 
Details of the model are presented in~\cite{rimoli2017reduced}.

To analyze the linear wave propagation response, we conduct a  
dispersion analysis on a single unit cell of the periodic tensegrity lattice. 
Periodic boundary conditions are imposed relating the displacements on the nodes at the two opposite square facets 
at the boundary of the unit cell, illustrated by the same color in Fig.~\ref{Tlattice}. 

All the calculations in this letter are done
with material properties for bars and cables corresponding to titanium with density $4480$ $\text{kg/m}^3$ and Young's modulus 
$100$ GPa. The bar and cable diameters are taken to be 2.3 cm and 1.15 cm, respectively, while the length of the
canonical unit cell is $1.26$ m. 
The frequencies are normalized by a reference frequency $\omega_r = c_0/a$, where $a$ is the magnitude
of the lattice vector and $c_0=\sqrt{E/\rho}$ is the longitudinal wave speed in the bulk material comprising the tensegrity structure. 
The non-dimensional frequency is thus given by $\Omega = \omega / \omega_r$. 
This choice of normalization ensures that the dispersion diagrams are valid for tensegrity lattice made of any 
linear elastic material and over any length scale where linear elasticity assumptions are valid, provided that the 
relative geometric dimensions of the cable and bar diameters with respect to the unit cell length $a$  are the same. 
Although we study only one 
geometric configuration, we believe these results are representative of a wide range of cable and bar dimensions.  

\section{Dispersion analysis of $1D$, $2D$ and $3D$ tensegrity media}\label{dispSec}
The dispersion analyses of tensegrity based beams, plates and solids are presented for different levels of cable prestrain $\lambda$. 
We examine the low frequency (long wavelength) propagation
of acoustic modes through the lattice and contrast their behavior with waves in conventional bulk media. 
Note that due to the presence of free surfaces (faces) 
in tensegrity beams and plates, the strain in the various cables (and thus bars), lying in the interior or on a free surface, 
may be different in the reference configuration. 
As the prestrain increases (decrease in $\lambda$), the bars experience increasing compression and eventually all the bars buckle. 
To illustrate qualitative differences between the cases when the bars are unbuckled and
buckled in the equilibrium configuration, we present dispersion diagrams for two prestrain levels $\lambda$: 
one where bars are unbuckled ($\lambda = \lambda_u=0.993$) and another
where all the bars buckle ($\lambda = \lambda_b=0.988$). 
It is worth emphasizing that a post-buckled dynamic analysis is possible due to the unique ability of our 
lattice to be stable and support loads even after the bars buckle~\cite{rimoli2017mechanical}. 

Let us make a technical remark on the mechanics of a buckled bar. 
In each bar, there is a $U(1)$ gauge freedom in the choice
of the buckling plane, i.e., the transverse displacement of a buckled bar can be in any direction orthogonal to its axis. 
Let us consider a local coordinate system 
with the $x$-axis along the centerline and the $y$-axis along the transverse displacement direction. 
The quasistatic response of the post-buckled beam (and hence lattice) 
is gauge independent as the transverse ($y$ direction) force (and stiffness) in the plane
of buckling is the same as that normal to this plane (i.e.,along $z$). However, under dynamic excitation, 
this gauge invariance is broken
as the force in the transverse direction ($y$) is nonzero while it is zero in the out of plane ($z$) direction. 
Thus our post-buckled results will depend on the choice of our gauge for the buckled configuration 
of each bar. Our numerical simulations showed that this choice does not
significantly influence the dispersion results 
at low frequencies and long wavelengths (wavevector $\bkappa \to \bzero$). At large 
wavevectors $\kappa$ and high frequencies, there are small variations in the dispersion behavior. 
In this work, we choose a gauge such that the beams always buckle in the local $xy$ plane, with $x$ being along the 
direction $\bv$ of the line joining the two end points and $y$ given by the unit vector along $\be_3 \times \bv$, with $\be_3$
being the unit vector along the global $z$ direction. 

\subsection{$1D$ lattice: Tensegrity beams}

\begin{figure*}
\centering
\subfigure[]{
\includegraphics[width=0.31\textwidth]{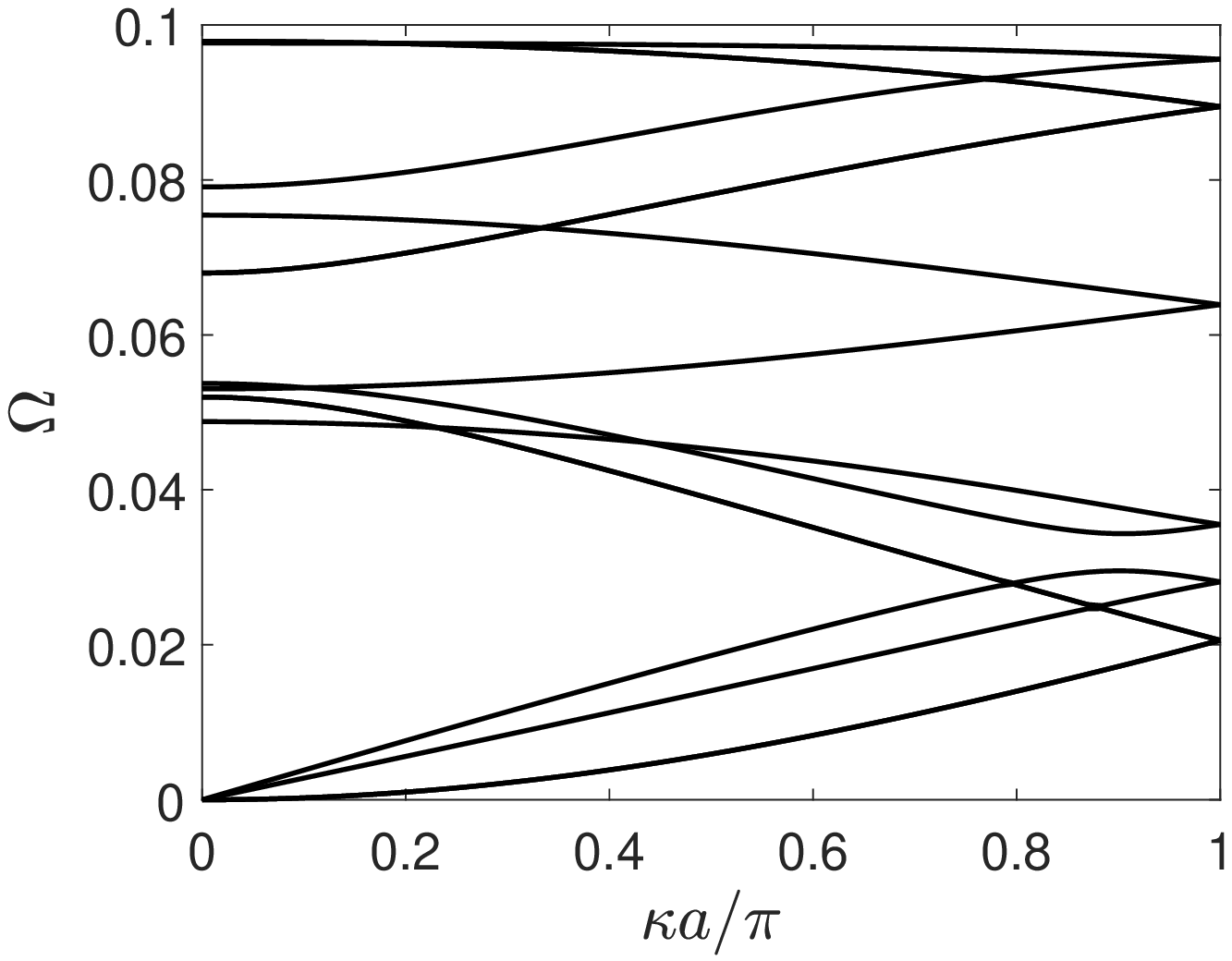}
\label{Disp_1D_993}
}
\subfigure[]{
\includegraphics[width=0.31\textwidth]{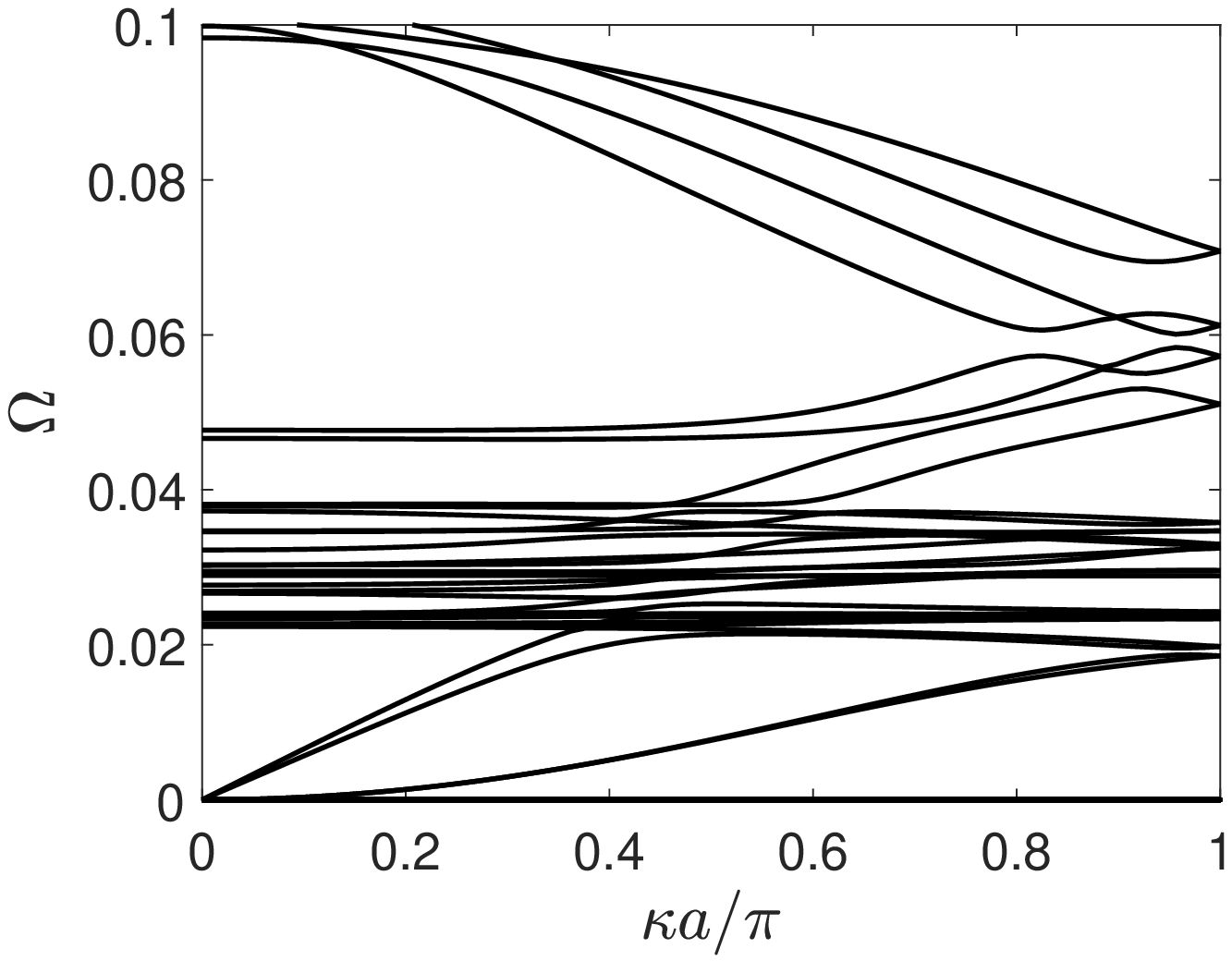}
\label{Disp_1D_990}
}
\subfigure[]{
\includegraphics[width=0.31\textwidth]{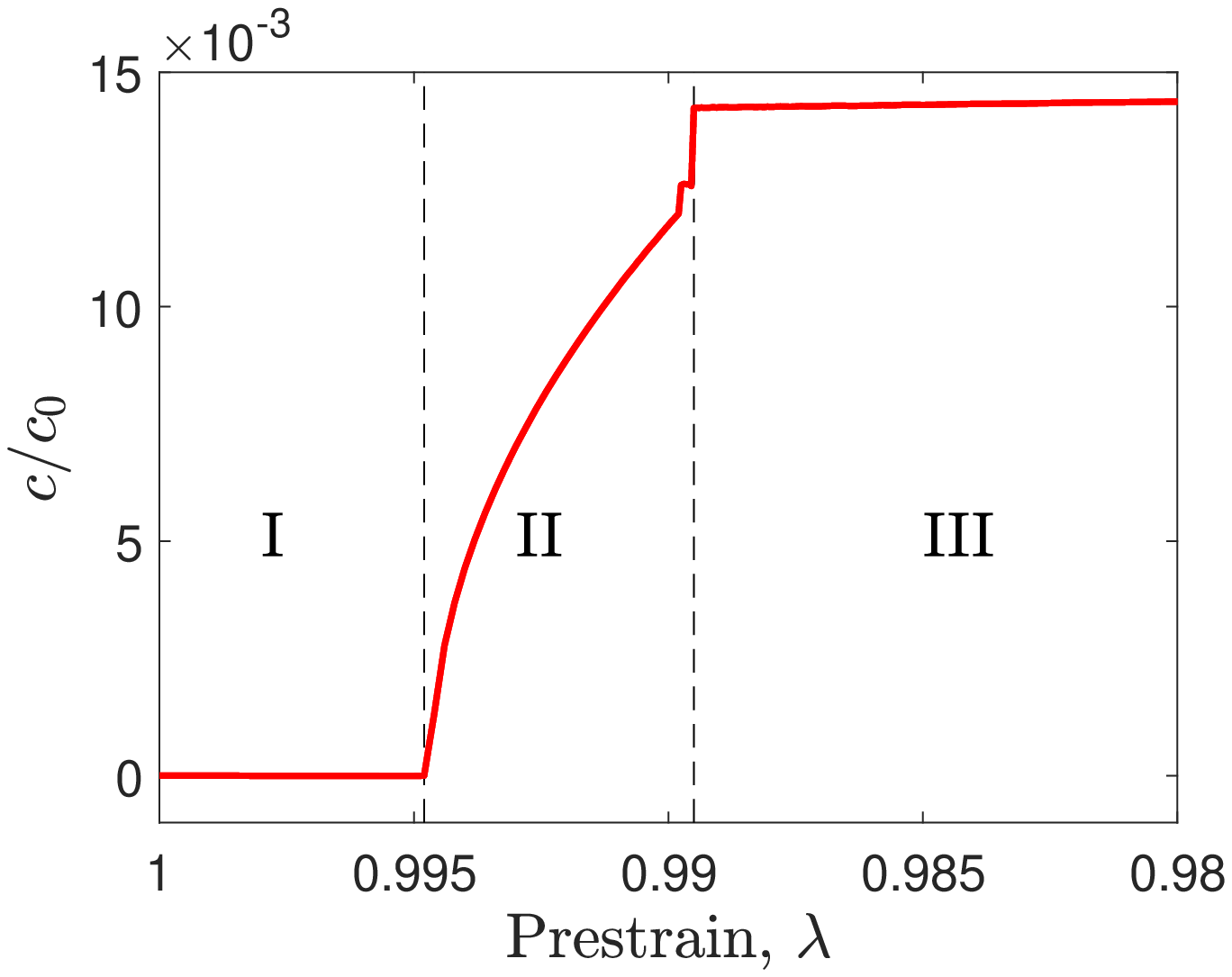}
\label{WaveSpeed_1D}
}
\caption{Tensegrity beam dispersion diagram for bars (a) prebuckled ($\lambda_u$) and (b) post-buckled ($\lambda_b$), 
with an axial, a torsional and two bending modes at low $\Omega$. The flat branches are due to soft modes 
in the buckled configuration. 
(c) Effect of prestrain on the wave speed of various waves showing two sharp transitions: when the lattice becomes stiff
and when the bars buckle.  } 
\label{Disp_1D}
\end{figure*}
The tensegrity beam is obtained by tesselating the unit cell in Fig.~\ref{Tbeam} along the $x$ direction. 
Its unit cell comprises of $24$ bars and $68$ cables. 
Figure~\ref{Disp_1D} displays its dispersion diagram for the two prestrain levels discussed above. 
For a prestrain $\lambda_u$ (Fig.~\ref{Disp_1D_993}), under which the 
bars are in compression, while all the cables are in tension, there are four acoustic modes: two coincident bending modes, 
an axial mode and a torsional  mode, in increasing order of frequency. 
The coinciding of the two bending modes follows from their orthogonality and from 
the $C_4$ symmetry of the beam unit cell about its axis. 

Figure~\ref{Disp_1D_990} 
displays the dispersion diagram for a cables prestrain $\lambda_b$, where the bars are buckled. 
In addition to the four acoustic modes, it has a collection of $24$ flat modes at low frequencies 
($\Omega \sim 0.02-0.04$). The bucking of the bar introduces these low frequency wave modes
and these modes are defined as soft modes. There are also $24$ zero frequency modes $(\Omega = 0)$
in the dispersion diagram. 
Each buckled bar contributes to a zero energy mode and a soft mode and these additional modes 
arise as a consequence of buckling in a single bar. 
A buckled bar can freely rotate about the line joining 
its ends since the pin joints do not offer any resistance. 
This rotation about the axis is a zero energy (floppy) mode and leads to the  zero frequency branch in the dispersion diagram. 
The flat modes at finite frequency in Fig.~\ref{Disp_1D_990} arise from the interaction of the cables with the 
axial deformation of the buckled bars. Since the axial stiffness of a buckled beam is significantly lower, this mode has a low
frequency. 

Having demonstrated qualitative differences between tensegrity beams with bars in 
the pre- and post-buckled configurations, let us now quantify this difference by analyzing the 
wave speeds in the long wavelength regime ($\bkappa \to \bzero$). These are computed by using the 
approximation $c = \partial \Omega / \partial \kappa \approx \Delta \Omega/\Delta \kappa = \Omega/\kappa$ at a $\kappa$ close to zero.   
The wave speeds are normalized by $c_0 = \sqrt{E/\rho}$. 
Figure~\ref{WaveSpeed_1D} displays the axial acoustic mode wave speed 
as a function of the prestrain level $\lambda$. Three distinct regimes are observed: 
at low prestrains (labeled I in the figure), the
wave speeds are zero. There are floppy modes (mechanisms) which result in these 
zero energy (or zero frequency) modes. At moderate prestrain values (labeled II in the figure), the wave speeds increase 
 monotonically from zero. The bending wave speed is also nonzero, but it is small compared to the axial wave speed. 
At a prestrain level $\lambda = 0.9898$, $8$ bars in the unit cell buckle, resulting in a jump in the wave speed. At $\lambda = 0.9895$, 
the remaining $16$ bars buckle, resulting in a second, larger jump in wave speed. 
Finally, as the prestrain increases further, the axial wave speed remains at a near constant value. 
The transitions between the stages are analogous to phase transitions in condensed matter systems~\cite{jaeger1998ehrenfest}. 
These results illustrate how the wave speeds of our tensegrity-based 
beams can be varied over a wide range by varying the cable prestrain.

\subsection{$2D$ lattice: Tensegrity plates}

\begin{figure*}
\centering
\subfigure[]{
\includegraphics[width=0.31\textwidth]{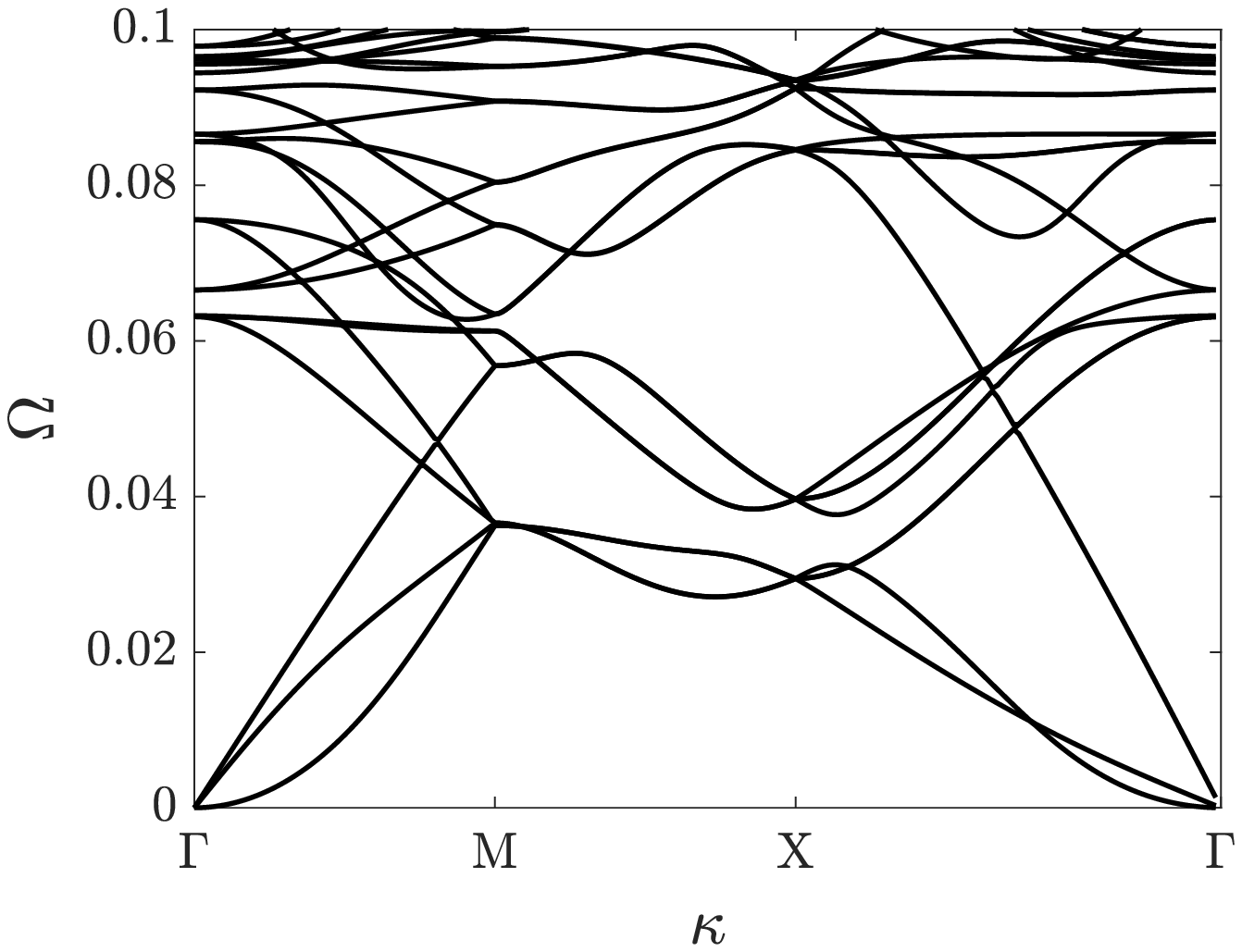}
\label{plate_disp_993}
}
\subfigure[]{
\includegraphics[width=0.31\textwidth]{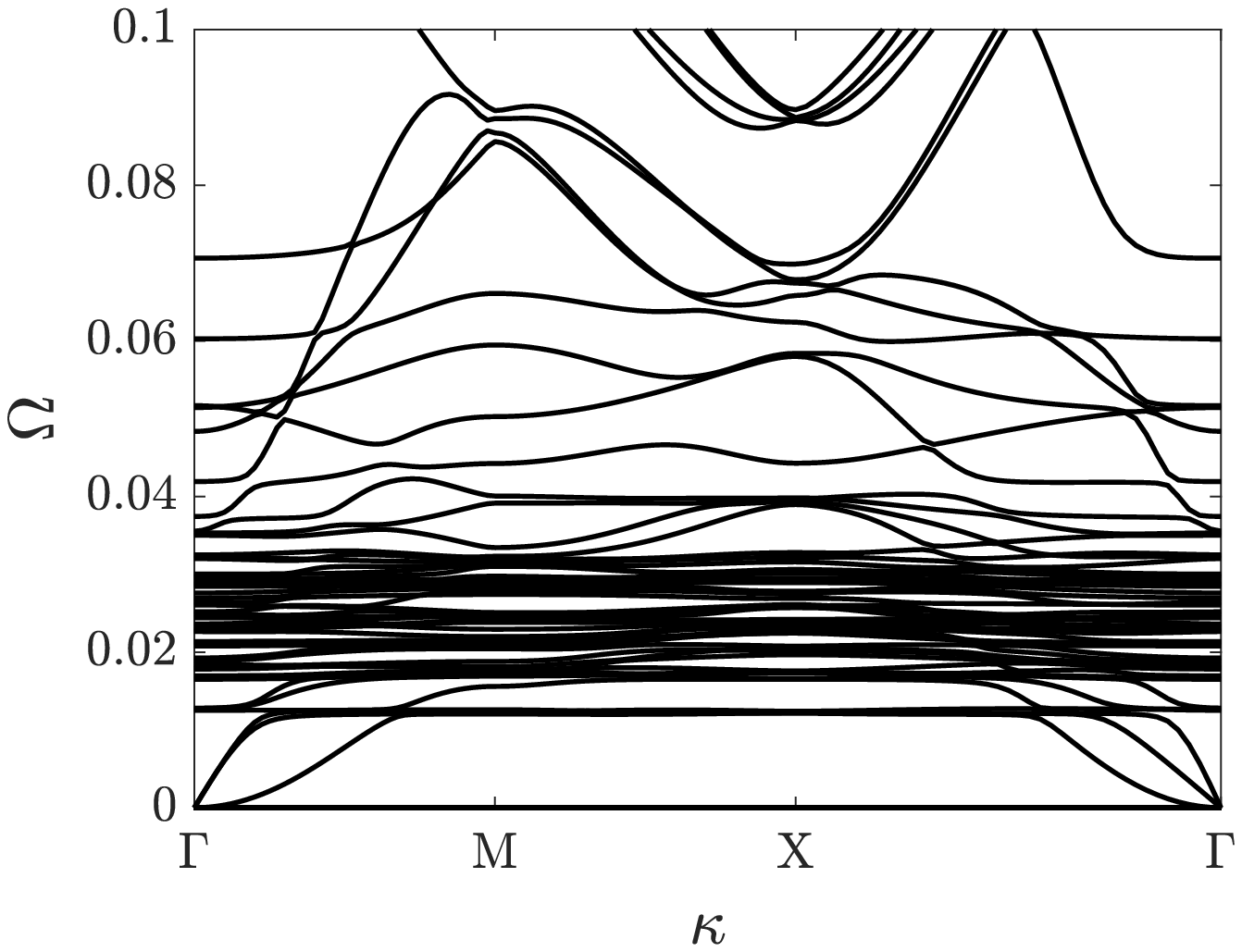}
\label{plate_disp_990}
}
\subfigure[]{
\includegraphics[width=0.31\textwidth]{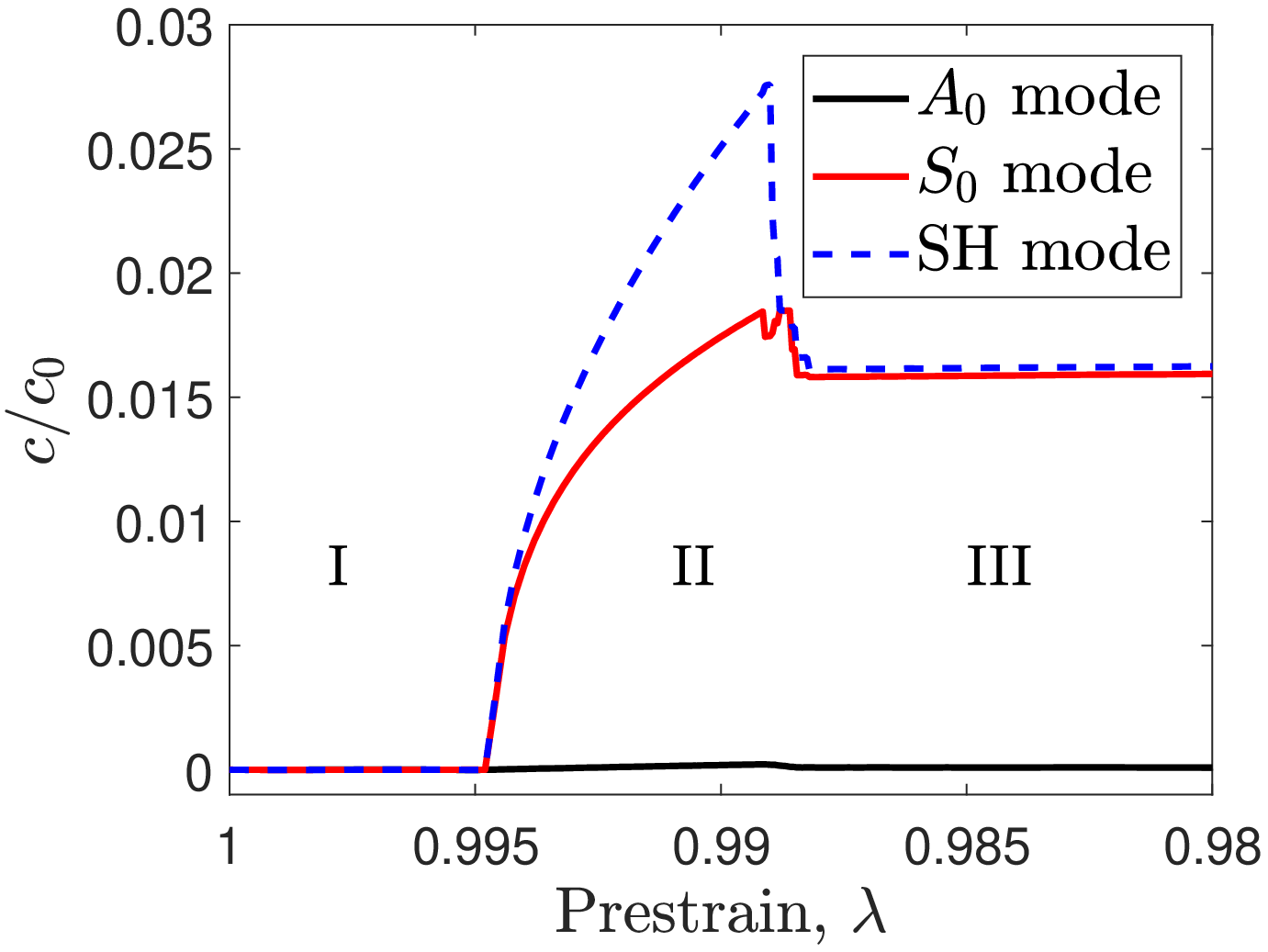}
\label{plate_wave speed}
}
\caption{Tensegrity plate dispersion diagram along the IBZ corners for (a) prebuckled bars ($\lambda_u$) and 
(b) post-buckled bars ($\lambda = 0.988$), showing a bending mode,  two coincident $S_0$ modes and a 
transverse shear (SH) mode. The additional lines in (b) are due to soft modes. (c) Wave speed with prestrain for the 
various wave modes having two sharp transitions. }
\label{2D_disp}
\end{figure*}

We now turn attention to analyzing the dynamic behavior of tensegrity plate lattices obtained by tesselating the unit cell in
Fig.~\ref{Tplate} along $x$ and $y$ directions. 
We present the dispersion diagrams for two levels of 
cable prestrain ($\lambda_u$ and $\lambda_b$) along the corners of the irreducible Brillouin zone (IBZ). 
The IBZ corners span the path $\Gamma M X \Gamma$ with $\Gamma=(0,0)$, $M=(\pi/a,0)$ and $X=(\pi/a,\pi/a)$. 
Figure~\ref{plate_disp_993} displays the low frequency modes for the 
case where the cable prestrain is set to $\lambda_u$. 
It has three acoustic modes, analogous to that observed in a plate made of a continuous solid. 

The first mode is analogous to an $A_0$ Lamb mode in a conventional plate
and is antisymmetric about the plate midplane. 
The second acoustic mode is symmetric about the plate midplane, 
analogous to the $S_0$ Lamb wavemode while the 
third acoustic mode has an inplane shear mode shape, analogous to 
$SH$ modes in plates. Similar to a conventional plate, the $A_0$-like mode has the lowest wave speed, followed by the $S_0$-like mode 
and the in-plane shear mode. 
Figure~\ref{plate_disp_990} displays the dispersion diagram for the lattice with cable prestrain 
set to $\lambda_b$. Similar to the $1D$ case, $48$ flat modes corresponding to zero energy modes are present at
$\Omega = 0 $ and around $\Omega = 0.02$.  
The acoustic mode shapes in the post-buckled case are identical to the presented case of a prestrain
level where the bars are unbuckled.

Let us analyze the effect of cable prestrain on the wave speeds to quantify the difference in behavior 
when the bars are pre- and post-buckled. 
Figure~\ref{plate_wave speed} displays the variation
of the acoustic mode wave speeds with prestrain parameter $\lambda$ along the lattice unit vector ($x$ or $y$) directions. 
Similar to the $1D$ case, three distinct
regimes are observed. At low prestrains (labeled I in the figure), 
the wave speeds are zero, while it monotonically increases from zero
at moderate prestrains. This range (labeled II in the figure) corresponds to the bars being unbuckled. 
As the bars begin to  buckle with increasing prestrain, the $S_0$ and $SH$ mode wave speeds decrease to an 
almost constant value. The bars begin to buckle in sets of $4$ or $8$ starting from a prestrain $\lambda=0.9892$, in six stages
until a prestrain level $\lambda = 0.9882$, when all the bars are buckled. These sets of buckling correspond to small jumps 
in the $S_0$ mode wave speed around $\lambda = 0.98$ in Fig.~\ref{plate_wave speed}. 

\begin{figure*}
\centering
\subfigure[]{
\includegraphics[width=0.3\textwidth]{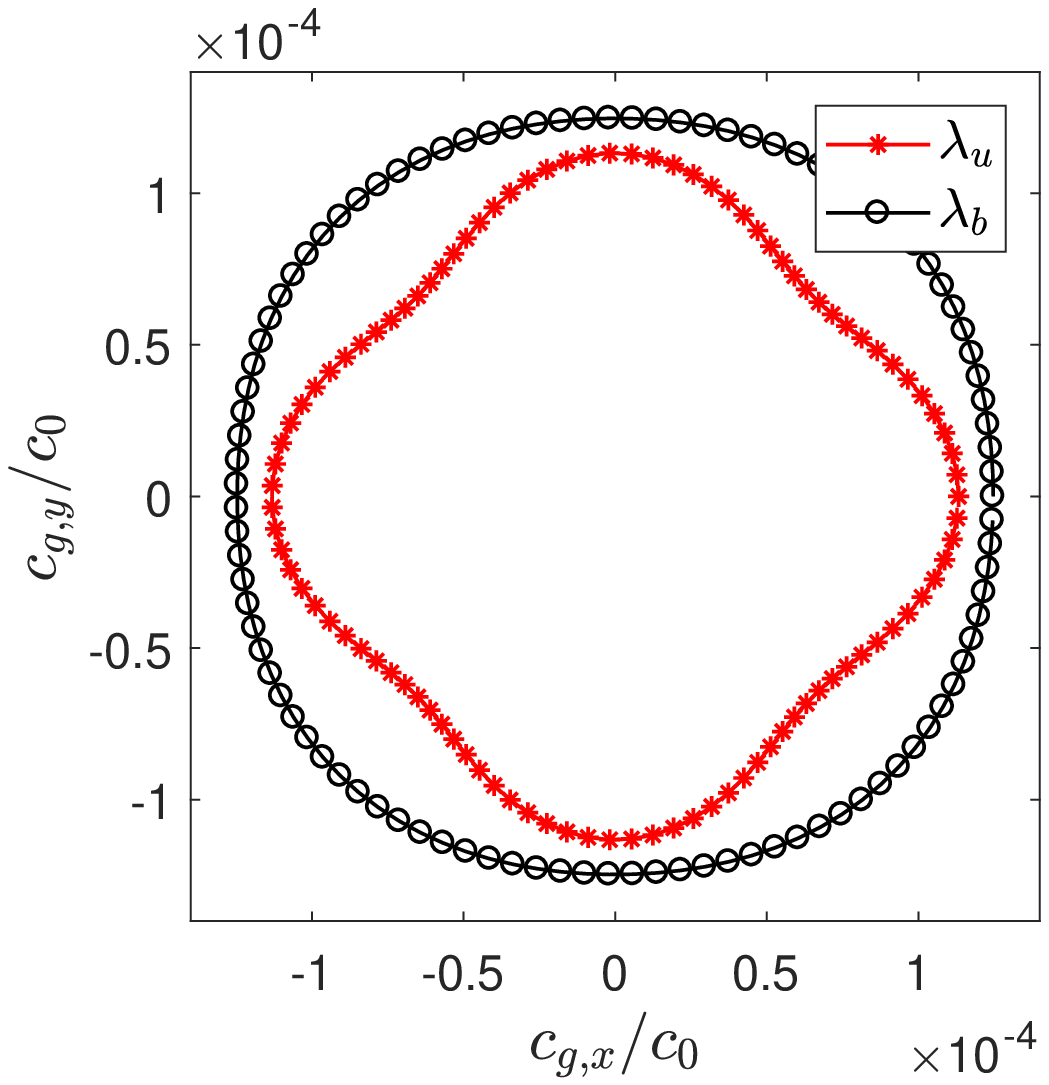}
\label{2D_combinedcontr}
}
\subfigure[]{
\includegraphics[width=0.3\textwidth]{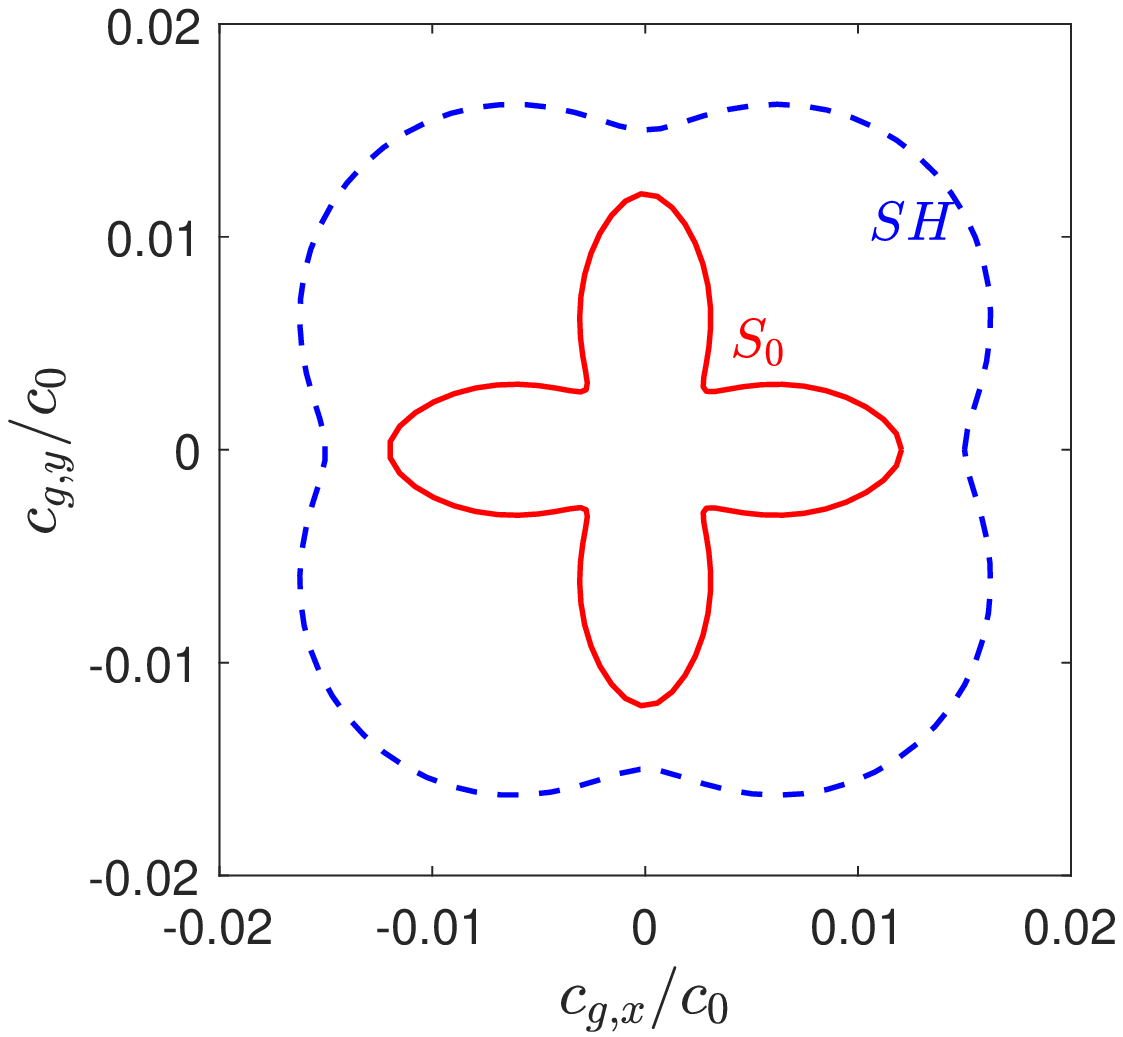}
\label{2D_993contr}
}
\subfigure[]{
\includegraphics[width=0.3\textwidth]{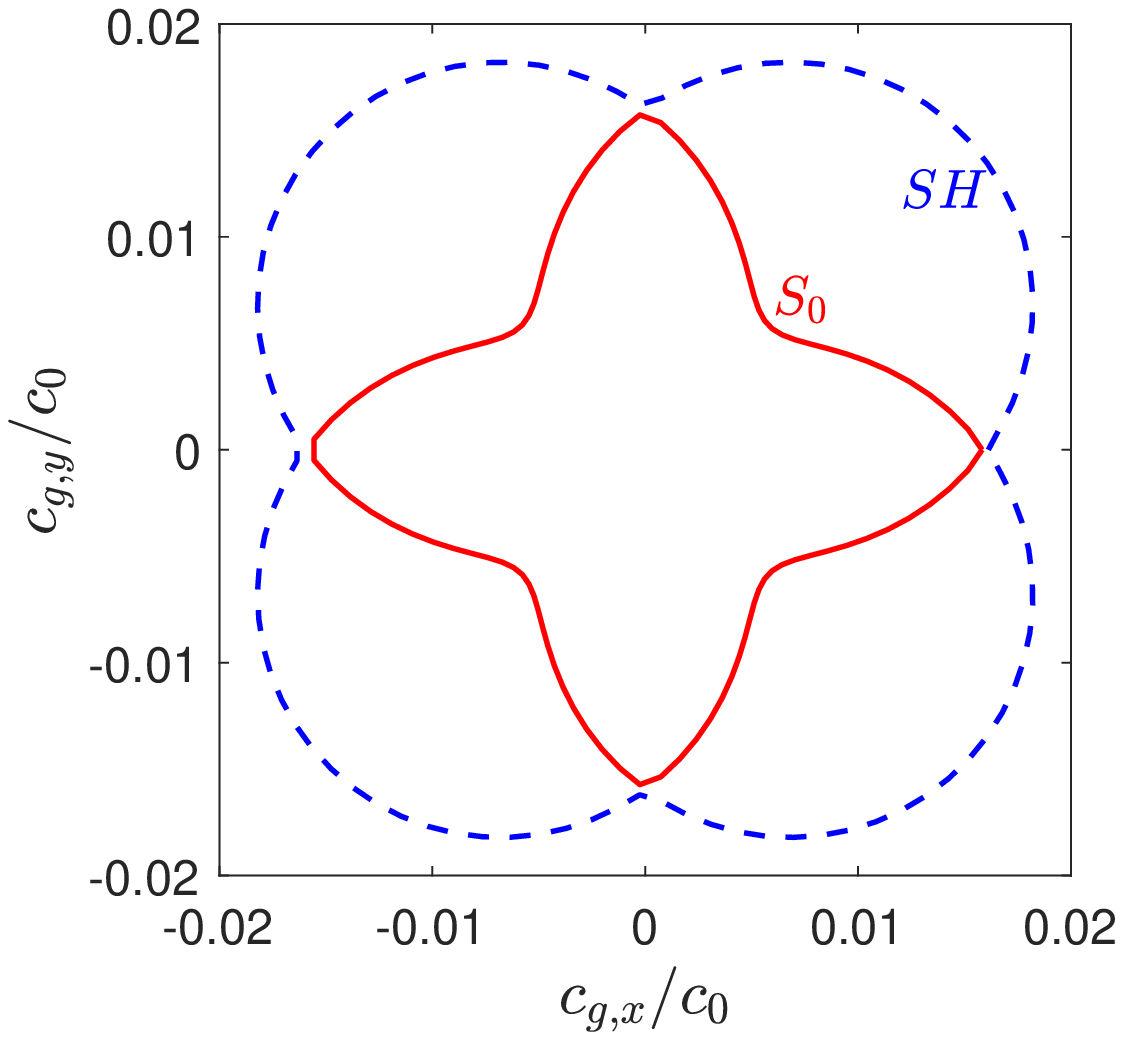}
\label{2D_990contr}
}
\caption{Wave speed variation with direction near $\bkappa = \bzero$ for two prestrain levels. 
(a) Flexural or $A_0$ mode. $S_0$ and $SH$ modes for bars in (b) pre-buckled and (c) post-buckled configuration. }
\label{2D_WaveSpeeds}
\end{figure*}

Let us now look at the wave speeds in all directions to understand anisotropy in the 
lattice response to a point excitation. 
For each wave-vector $\bkappa = (\kappa_x,\kappa_y)$ near $\bzero$ and mode $m$ with frequency $\Omega_m$, 
we compute the group velocity $\bc = (c_{g,x},c_{g,y})$ in each direction by approximating $\partial \Omega_m/\partial \bkappa$, i.e.,  the 
components of the group velocity are $c_{g,x} = \Omega_m / \kappa_x$ and $c_{g,y} = \Omega_m / \kappa_y$. 
The variation with direction is obtained by simply setting $\bkappa = \kappa (\cos\theta,\sin\theta)$ and 
spanning $\theta$ in $[0,2\pi]$. $\kappa$ is chosen to be sufficiently small to get the wave speeds at low frequency.  

Figure~\ref{2D_WaveSpeeds} displays the directional variation of wave speeds of the three
acoustic modes for the two prestrain levels, $\lambda_u$ and $\lambda_b$. These
two cases are representative of the behavior in the second and third stages in Fig.~\ref{plate_wave speed}. 
Figure~\ref{2D_combinedcontr} displays the group velocity components for the $A_0$ or flexural mode for both the
prestrain cases. The group velocity variation 
exhibits a four fold symmetry, consistent with the $D_4$ lattice symmetry and becomes more isotropic as the bars 
become buckled ($\lambda_b$). 
Note that the group velocity decreases linearly to zero with decreasing $\Omega$, since the frequency 
scales as $\Omega \sim |\kappa|^2$ for the flexural mode.
Hence, this group velocity plot illustrates only the shape of the variation with direction 
as $\kappa \to 0$ (the magnitude of $c_g$ is meaningless here). 

Figure~\ref{2D_993contr} displays the 
group velocity variation with direction for the  $S_0$ and $SH$ modes at a prestrain level $\lambda_u$. 
The $S_0$ mode exhibits strong anisotropy 
and the group velocity is high along the lattice directions. In contrast the $SH$ mode is relatively isotropic and its maximum
wave speed is along the diagonal directions. Finally, Fig.~\ref{2D_990contr} displays the group velocity contours 
for the lattice with prestrain level $\lambda_b$. It is qualitatively similar to the unbuckled case, 
with a reduced anisotropy in the $S_0$ mode. 
In both the cases, the $S_0$ waves travel slower than the $SH$ waves in all directions, similar to the typical 
response of a plate made of conventional solids.

\subsection{$3D$ lattice: Tensegrity solids}
\begin{figure*}
\centering
\subfigure[]{
\includegraphics[width=0.45\textwidth]{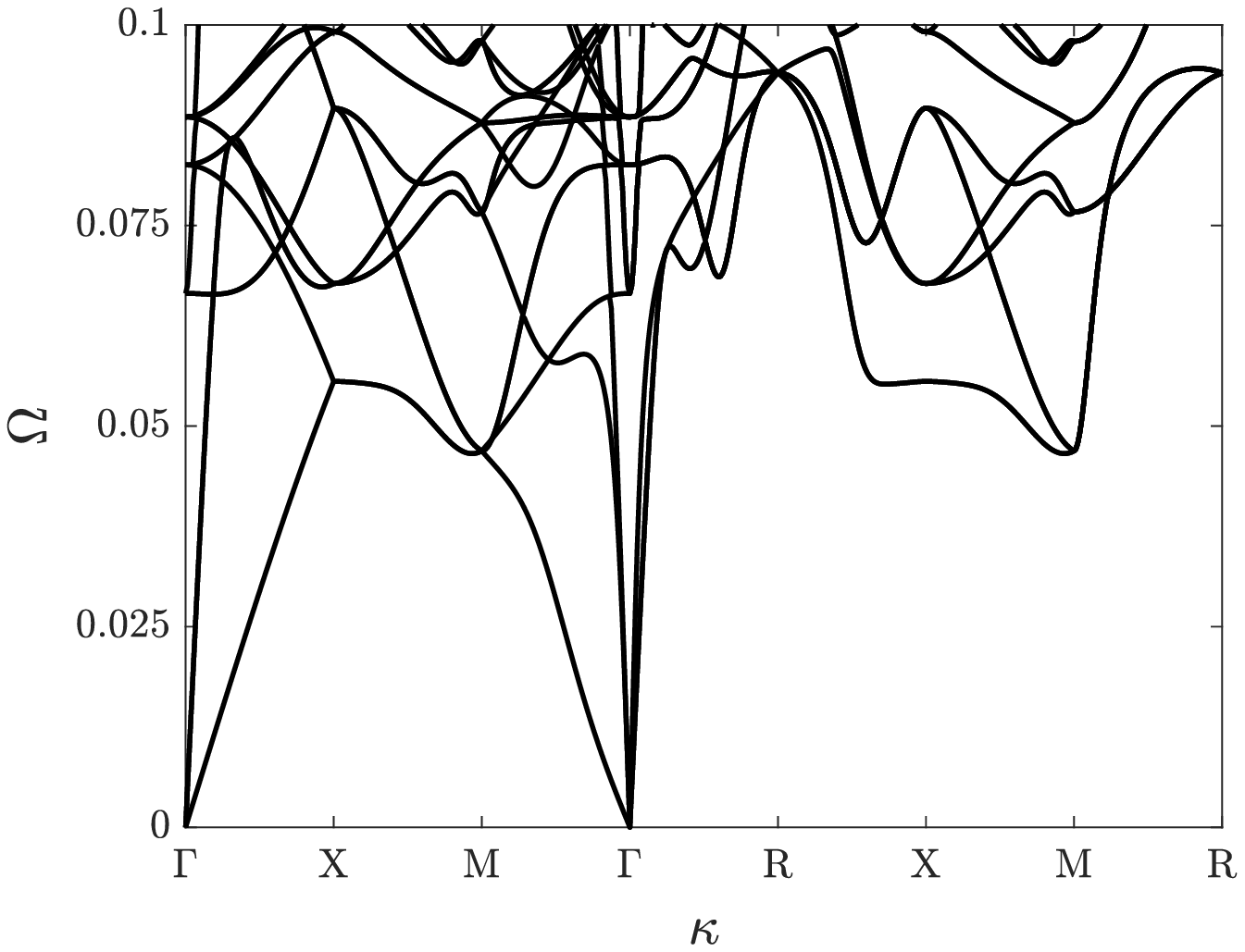}
\label{solid_disp_993}
}
\subfigure[]{
\includegraphics[width=0.45\textwidth]{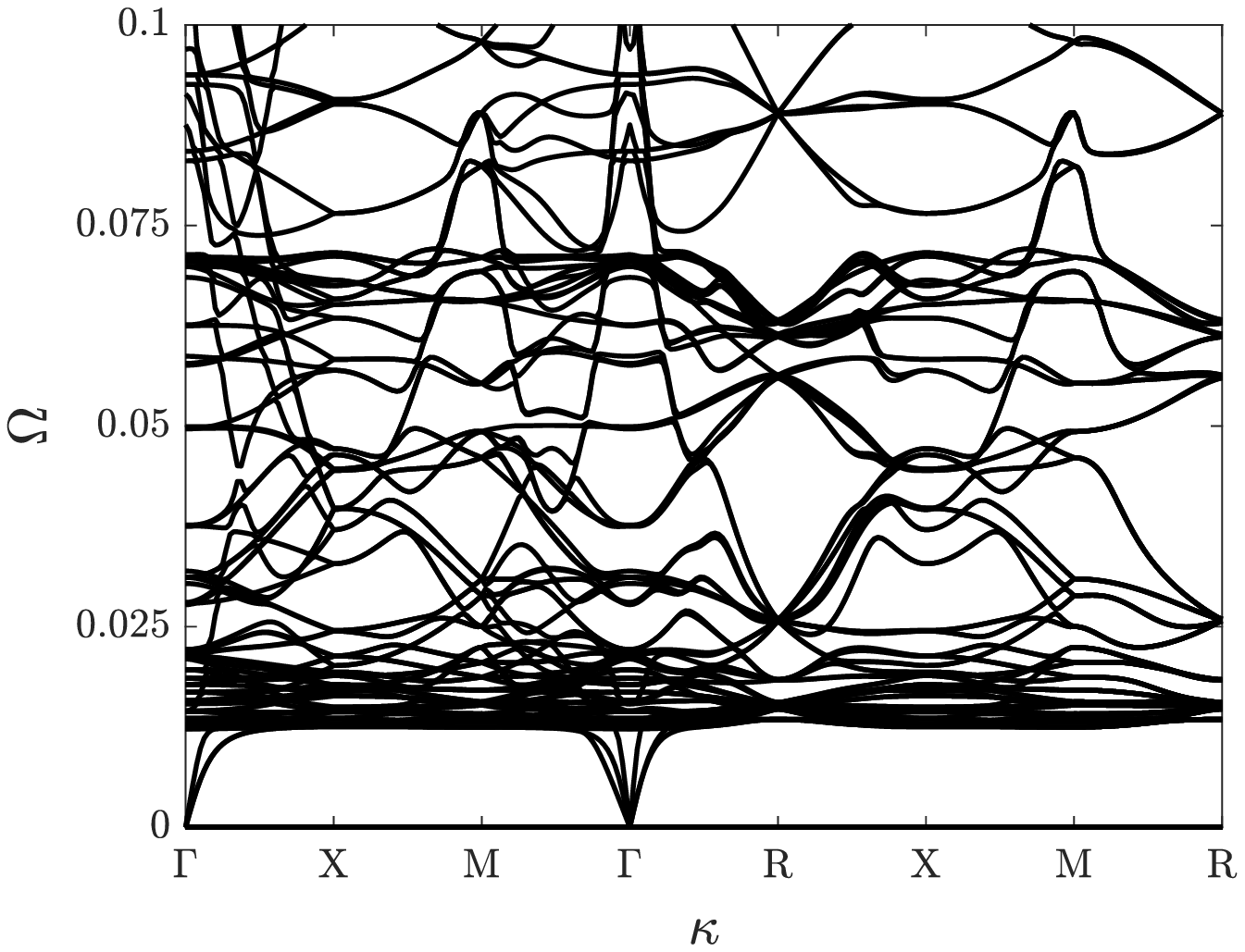}
\label{solid_disp_990}
}\\
\subfigure[]{
\includegraphics[width=0.42\textwidth]{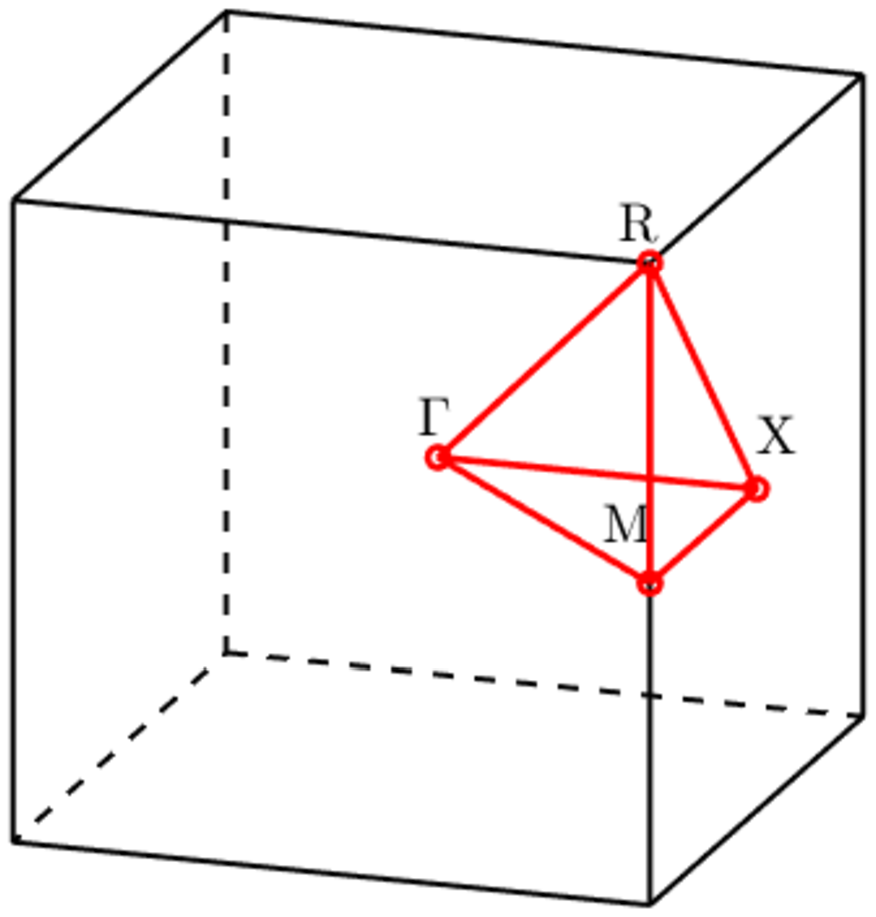}
\label{IBZ_3D}
}
\subfigure[]{
\includegraphics[width=0.48\textwidth]{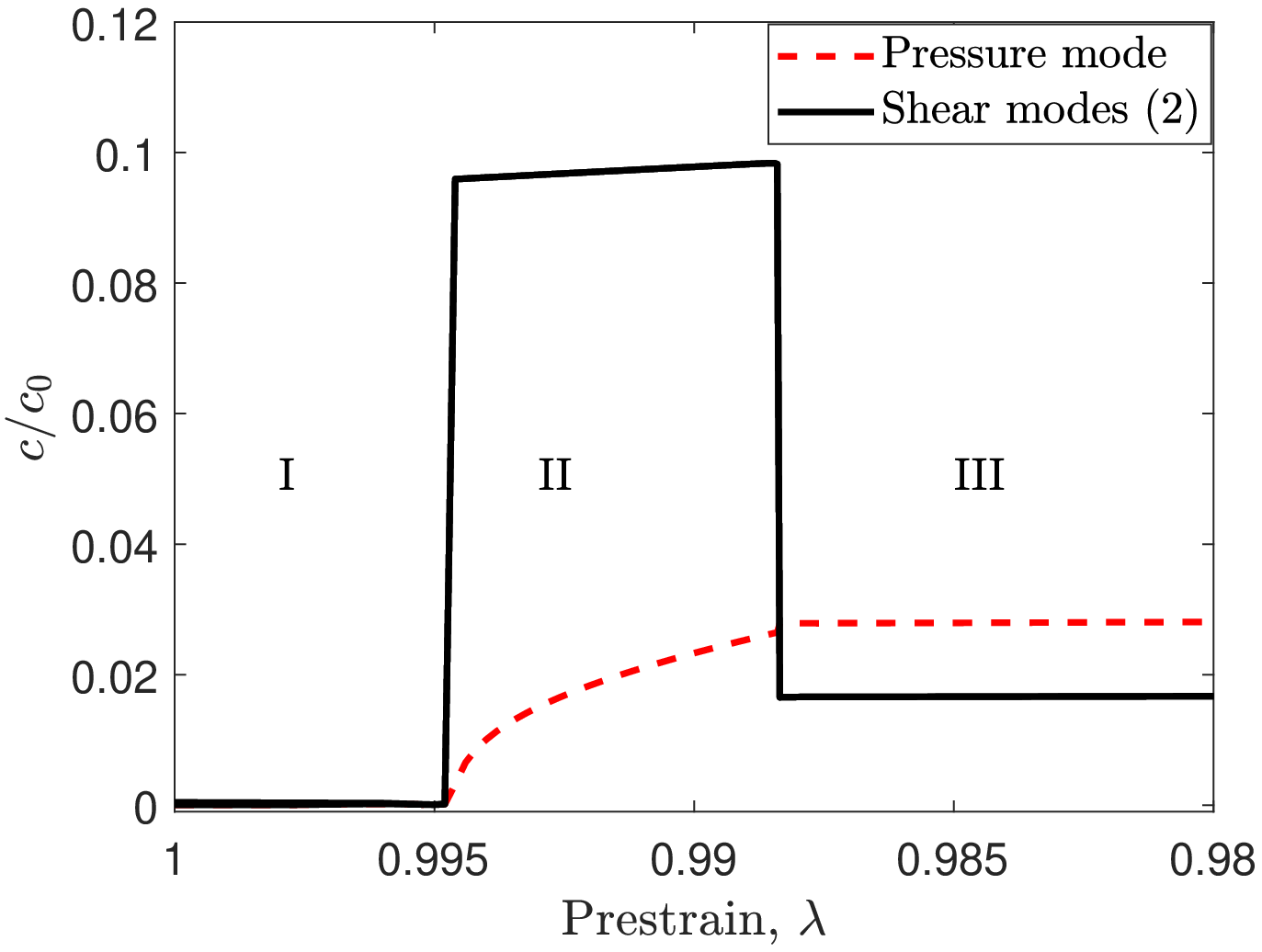}
\label{solid_wave speed}
}
\caption{Tensegrity solid dispersion diagram along the corners of the IBZ (shown in (c)) for  
(a) prebuckled bars and (b) post-buckled bars, both having one pressure mode and two shear modes at low $\Omega$. 
The flat curves in (b) again correspond to soft modes. (d) Effect of prestrain on the wave speeds. Shear waves 
travel faster than pressure waves in the prebuckled configuration, while this gets reversed in the post-buckled configuration. 
} 
\label{3D_disp}
\end{figure*}

Let us now analyze the dynamic behavior of tensegrity solids obtained by tesselating its unit cell (Fig~\ref{Tsolid}) along 
$x$, $y$ and $z$ directions.  
The lattice unit cell has cubic symmetry as all the cables have identical prestrain~\cite{amirPaper}. The 
first Brillouin zone is a cube and the irreducible Brillouin zone is a tetrahedron~\cite{bradley2010mathematical}, 
illustrated in Fig.~\ref{IBZ_3D}.  
Figure~\ref{solid_disp_993} displays the first few modes of the dispersion diagram for a prestrain $\lambda_u$. 
There are three acoustic modes: a pressure mode and two shear modes. 
The shear modes are doubly degenerate along the $\Gamma X$ direction as a consequence of the cubic symmetry of the lattice~\cite{amirPaper}. 
Contrary to most solids~\cite{achenbach2012wave}, the two shear waves have a higher
group velocity than the longitudinal wave in all directions. 
The contour diagrams of group velocity variation 
presented later in this section illuminate this behavior clearly. 
Figure~\ref{solid_disp_990} displays the dispersion diagram at low frequencies when the cables are subjected to 
a prestrain $\lambda_b$, where the bars are buckled in the equilibrium 
configuration. Similar to the dispersion diagrams of tensegrity beams and plates, there are $96$ zero frequency modes
and an equal number of soft modes. 
The acoustic shear mode gets hybridized with these soft modes and it does not
propagate in these band of frequencies. These flat bands effectively act as stop bands only for shear waves. 

Figure~\ref{solid_wave speed} displays the variation of wave speeds for the long wavelength acoustic modes along the $x$-direction 
as the cable prestrain is varied. Again, three distinct regions are observed, with sharp jumps between them, analogous to phase 
transitions in solids. 
As the prestrain parameter $\lambda$ decreases below 0.995, the
shear and pressure wave speeds jump from zero to a finite value. 
They remain almost constant for the entire range of prestrain where the bars are 
prebuckled. The shear wave speeds are higher than the pressure wave 
speed. As the cable prestrain increases, the bars buckle  at $\lambda = 0.9885$
and the wave speed decreases. Note that all the bars 
are under the same compression in the reference configuration due to symmetry and they all buckle 
simultaneously. When the bars buckle, the pressure wave speed becomes
higher than the shear wave speed, similar to the behavior of conventional solids~\cite{achenbach2012wave}

\begin{figure*}
\centering
\subfigure[]{
\includegraphics[scale=0.48]{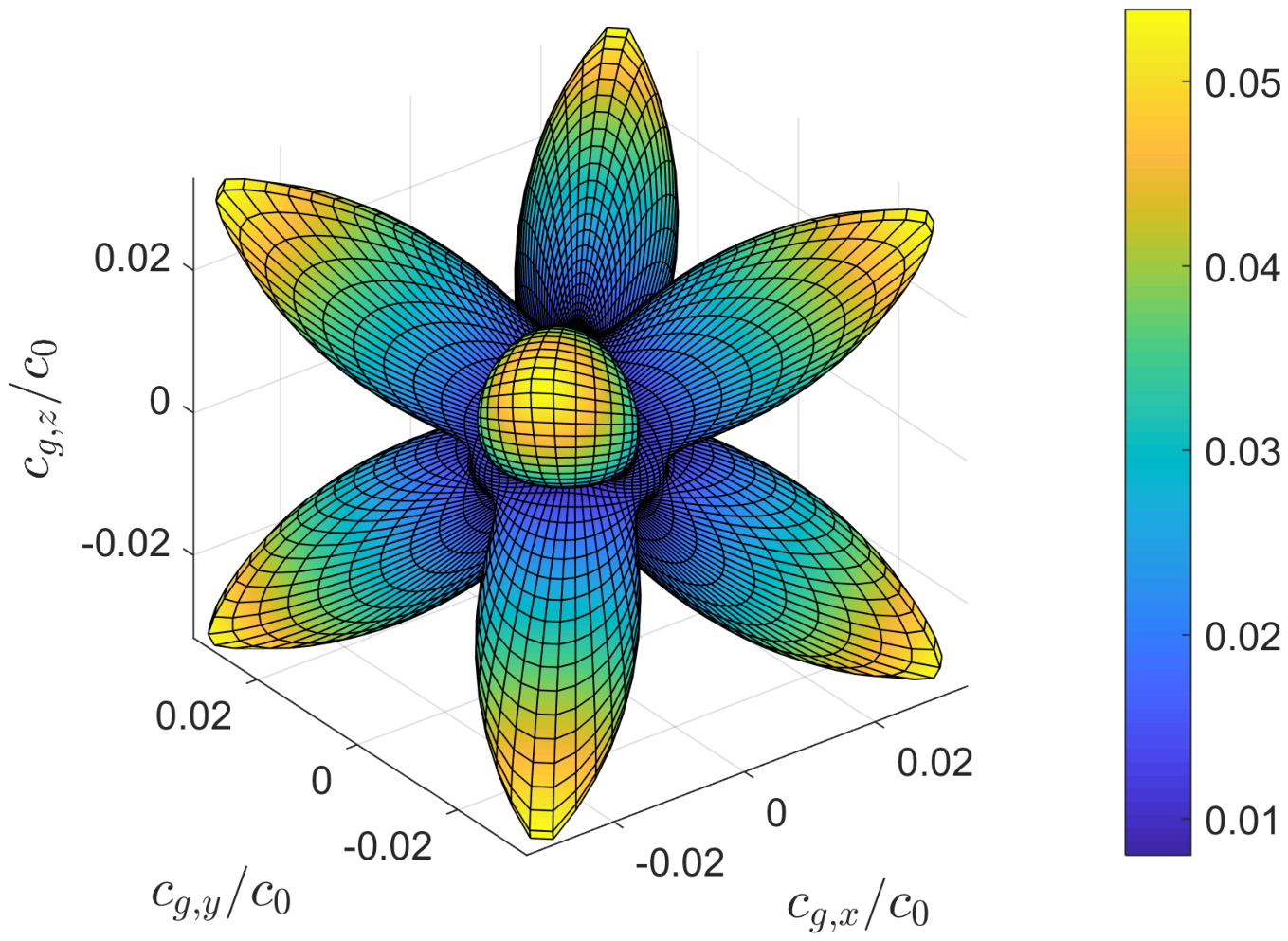}
\label{3d_pres_993_contr}}
\subfigure[]{
\includegraphics[scale=0.48]{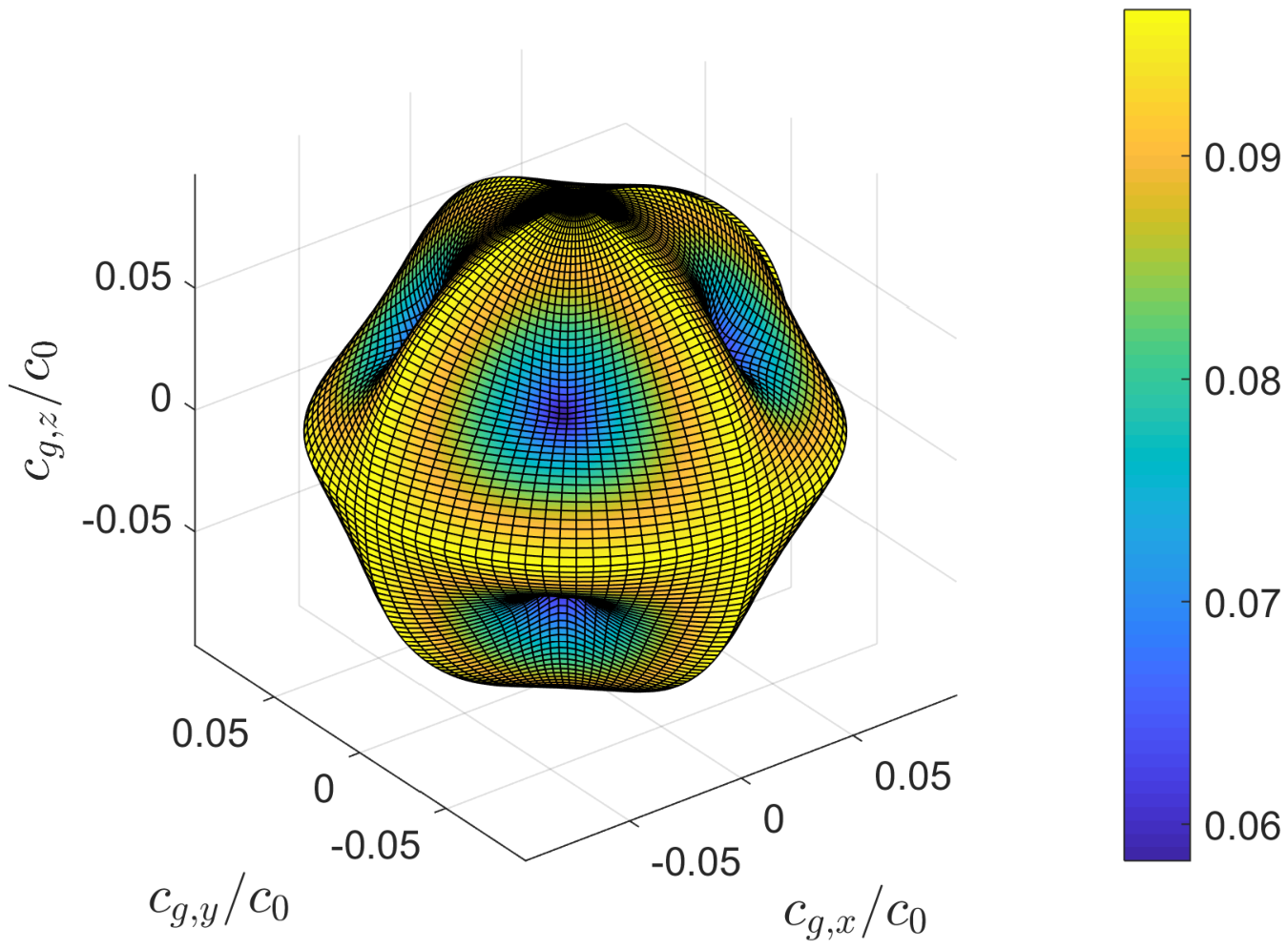}
\label{3d_pres_990_contr}}
\subfigure[]{
\includegraphics[scale=0.45]{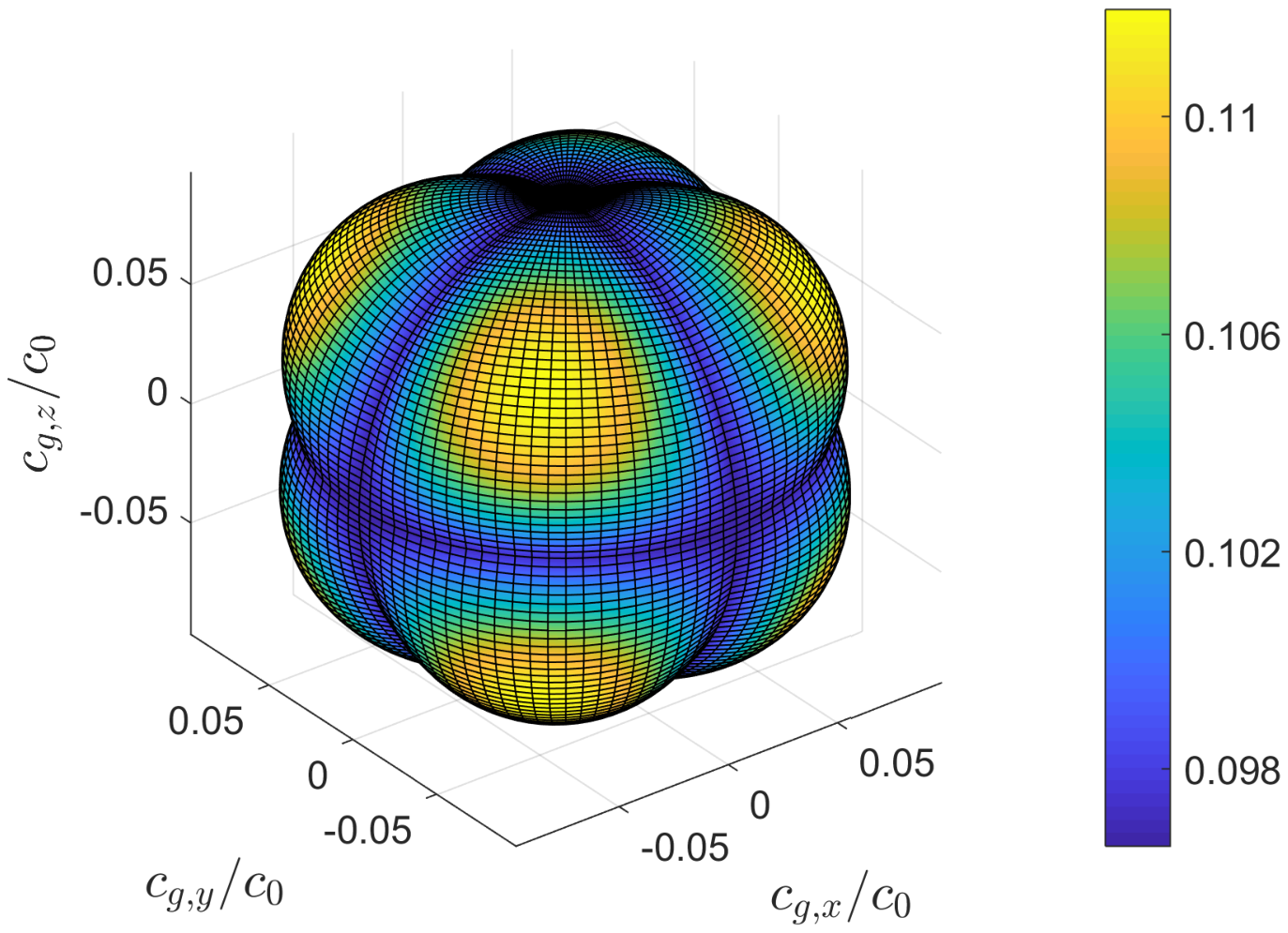}
\label{3d_shear1_993_contr}}
\subfigure[]{
\includegraphics[scale=0.45]{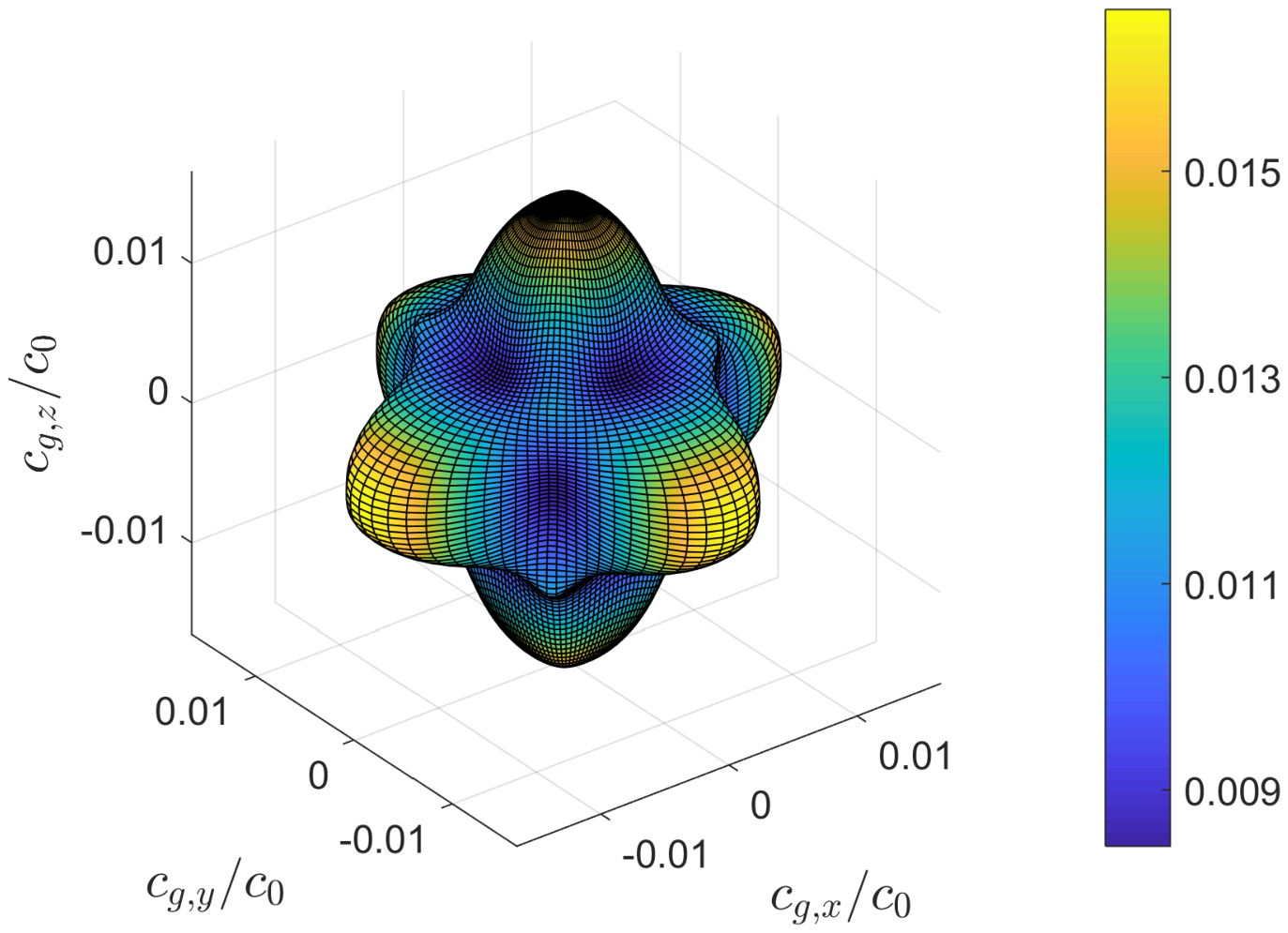}
\label{3d_shear1_990_contr}}
\subfigure[]{
\includegraphics[scale=0.45]{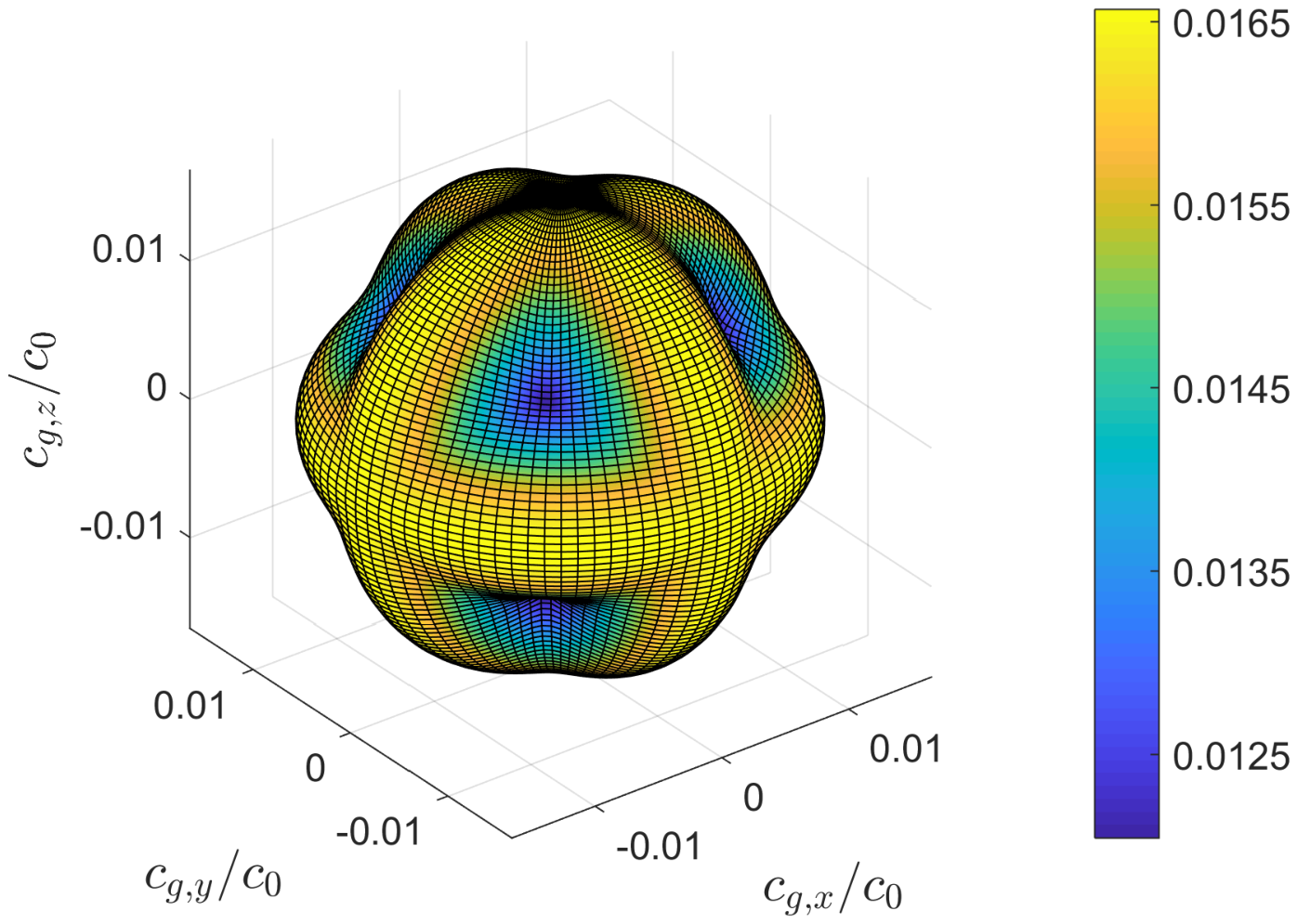}
\label{3d_shear2_993_contr}}
\subfigure[]{
\includegraphics[scale=0.45]{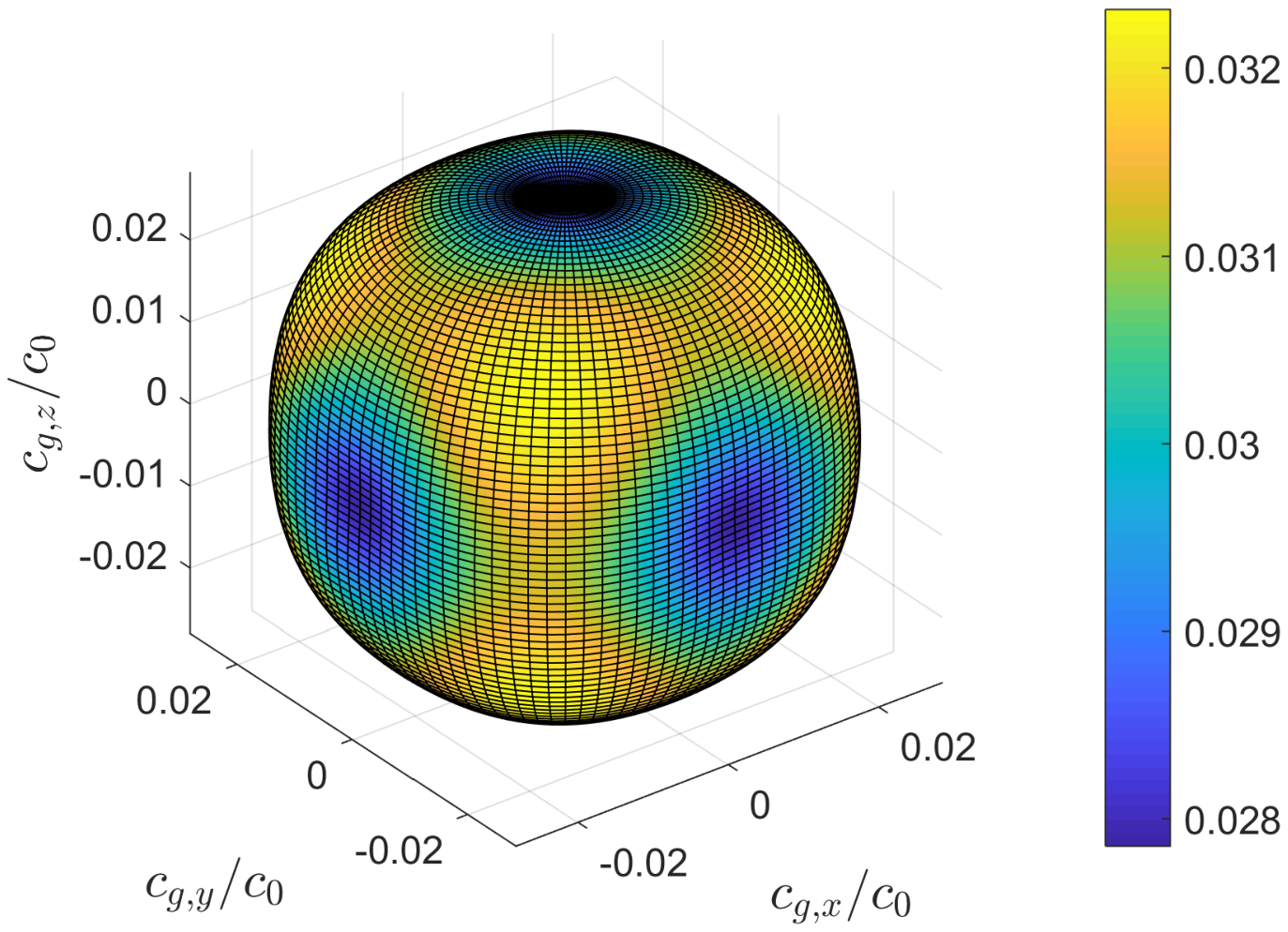}
\label{3d_shear2_990_contr}}
\caption{Dispersion surface of the tensegrity solid near $\bkappa=\bzero$ for the prebuckled (a,c,e) and 
post-buckled (b,d,f) configurations. 
(a) and (b) show pressure modes while the remaining plots show the two shear modes for each configuration. The behavior anisotropic and 
changes as the bars in the unit cell buckle under prestrain. 
} 
\label{3D_DispSurface}
\end{figure*}

To gain insight on the directional variation of the acoustic mode wave speeds in the lattice, we examine
the dispersion surfaces in the long wavelength limit. A similar procedure to that used for plates
is employed here for computing this directional variation. We span the directions associated with the set of normal vectors 
to a unit sphere by setting
$(\kappa_x,\kappa_y, \kappa_z) = \kappa(\cos\phi \sin\theta, \sin\phi \sin\theta, \cos\theta)$. Here $\theta$ and $\phi$
span $[0,\pi]$ and $[0,2\pi]$, respectively, and $\kappa$ is sufficiently small to get the long wavelength
non-dispersive wave speeds.  
Figure~\ref{3D_DispSurface} displays the variation with direction
for the three acoustic modes, with both the color and the magnitude in the radial direction denoting the 
wave speed. When the cable prestrain level is $\lambda_u$,  
the pressure wave speed is higher along the long (example: $[1,1,1]$) diagonal directions (Fig.~\ref{3d_pres_993_contr}). 
The two shear modes (Fig.~\ref{3d_shear1_993_contr},(e)) 
in the prebuckled configuration have a directional variation close to an isotropic sphere. 
The degeneracy in the two shear acoustic modes in Fig.~\ref{3D_disp} is along the 
high symmetry directions ($\Gamma X$, $\Gamma R$ and $\Gamma M$) only, 
as evident from the distinct shapes of their dispersion surfaces (Fig.~\ref{3D_DispSurface}(e,f)).
Note from the range of the color bars that the lowest shear wave speed is higher than the maximum pressure wave 
speed. Thus, both shear modes are faster than the pressure mode in all directions. Similar trends have recently been observed
in dilational metamaterials in~\cite{buckmann2014three}.  

Figures~\ref{3D_DispSurface}(b,d,f) display the group velocity variation with direction when the 
cable prestrain is $\lambda_b$. 
The pressure wave speed increases 
along the coordinate directions and decreases along the diagonal $[\pm 1, \pm 1, \pm 1]$ directions. Note also from the range of the
colorbar that the mean wave speed value is significantly lower than the unbuckled case. Similarly, the shear wave 
speeds also decrease by a factor of about $5$ although the contour shape remains qualitatively similar. 
The pressure waves travel faster than the shear waves along the coordinate directions, while the shear waves travel faster
along the long diagonal directions, illustrating the unique dynamic properties of our tensegrity solid. 

The three key aspects of dispersion analysis in tensegrity solids: faster shear than pressure waves 
at low prestrains, flipping to conventional behavior with faster pressure waves and the 
flat bands acting as stopbands for shear waves are verified by numerical simulations
on finite lattices. Details are presented in the accompanying Supplementary Materials. 
Thus the dynamic response of our tensegrity solids can change significantly as the bars buckle due to the cable prestrain.  

\section{Conclusions}\label{concSec}

Tensegrity-based periodic media offer unique opportunities for tunable wave propagation by varying the 
cable prestrain. In this letter, we presented a linear wave propagation analysis of 
tensegrity-based beams, plates 
and solids. Dispersion analysis is used to investigate the dynamic response 
of these structural  units and it shows a significant change in their  
long wavelength acoustic mode wave speeds as a function of the cable prestrain. 
Three distinct stages are identified: at low prestrain levels, 
the wave speeds are zero as the structure has zero stiffness; at moderate prestrain levels when the bars are under compression
but not buckled, the wave speed is nonzero and finally, at prestrain levels where the bars have buckled, the wave speed 
is distinct. Sharp transitions in wave speeds are observed between the three stages,  analogous to phase transitions in a solid. 
Our tensegrity solids are observed to support an unusual dynamic property of faster shear waves than
pressure waves in all directions in the second stage. 
In the post-buckled regime, additional zero frequency modes and soft modes having flat dispersion bands are observed in all the three 
structural units. 
A collection of these flat bands essentially acts as stop bands for shear excitations and opens new avenues 
in achieving  low frequency bandgaps. 
In summary, the unique post-buckling stability of these lattices allows for exploiting these sharp 
effective phase transitions to get qualitatively distinct dynamic behaviors as the bars buckle.

\bibliographystyle{unsrt}
\bibliography{paper}

\end{document}